\documentclass[aps,prl,floatfix,twocolumn,nofootinbib,amsmath,amssymb]{revtex4}
\usepackage{amsmath,amsthm,amssymb}
\usepackage{latexsym,psfrag}
\usepackage{graphicx,color}

\newcommand{\rd}{{\mathrm d}}

\begin{document}

\title{Resummation for $W$ and $Z$ production at large $p_T$}

\author{Thomas Becher\,$^a$, Christian Lorentzen\,$^a$ and Matthew D. Schwartz\,$^b$}

\affiliation{$^a$\,Institut f\"ur Theoretische Physik, Universit\"at Bern\\
Sidlerstrasse 5, CH--3012 Bern, Switzerland\\
$^b$\,Department of Physics, Harvard University,Cambridge, Massachusetts 02138, USA}
\begin{abstract}
Soft-Collinear Effective theory is used to perform threshold resummation for $W$ and $Z$ production at large transverse momentum
 to next-to-next-to-leading logarithmic accuracy including matching to next-to-leading fixed-order results. 
The results agree very well with data from the Tevatron, and predictions are made for the high-$p_T$ spectra at the LHC.
While the higher-log terms are of moderate size, their inclusion leads to a substantial reduction of the perturbative uncertainty.
With these improvements, the PDF uncertainties now dominate the error on the predicted cross section.
\end{abstract}
\maketitle


The production of an electroweak boson at large transverse momentum is arguably the most basic hard-scattering process at hadron colliders. In fact, it was one of the first  for which the next-to-leading order (NLO) perturbative corrections were computed \cite{Ellis:1981hk,Arnold:1988dp,Gonsalves:1989ar}. 
By now the complete $\alpha_s^2$ corrections to vector boson production are known, but since the $p_T$-spectrum starts at $O(\alpha_s)$ its theoretical accuracy has not been improved since the nineteen eighties. Given that $Z$'s and $W$'s at high transverse momentum provide an important background to new physics searches and a standard way to calibrate jet energy scales, it is important to have good theoretical control of their cross sections.

In the absence of a full NNLO computation, a way to improve  predictions is to compute the contributions to the cross section which arise near the partonic threshold from the emission of soft and collinear gluons. For very large $p_T$, these corrections become dominant and must be resummed to all orders in perturbation theory. However, even away from this region, the corresponding terms yield in many cases a good approximation to the full cross section. For vector boson production, this resummation has been performed to next-to-leading logarithmic (NLL) accuracy in \cite{Kidonakis:1999ur,Kidonakis:2003xm,Gonsalves:2005ng}. 

In the present paper, we use Soft-Collinear Effective Theory (SCET)  \cite{Bauer:2000yr,Bauer:2001yt,Beneke:2002ph} to perform the resummation of the threshold terms to NNLL accuracy.  A detailed derivation of the factorization theorem and phenomenological analysis of the closely related process of direct photon production were presented by the authors in~\cite{Becher:2009th}. This paper is a generalization of those results to include vector boson masses which significantly complicates the calculations.
 
In the partonic threshold region, where the final state has an invariant mass much lower than the transverse momentum,
 the cross section for a given partonic initial state $I$ to produce an electroweak boson $V$ factorizes as
\begin{multline}\label{factform}
\frac{\rd\hat{\sigma}_{I} }{\rd\hat{s}\, \rd\hat{t}} = \hat{\sigma}^B_{I}( \hat{s},\hat{t})\, 
H_{I}(\hat{s},\hat{t},M_V,\mu)\\
 \times \int\! \rd k\, J_{I}(m_X^2-2 E_J k) S_{I}(k,\mu)\, ,
\end{multline} 
where the partonic Mandelstam variables are $\hat{s} = (p_1+p_2)^2$ and $\hat{t} = (p_1-q)^2$, with $q$ the vector boson momentum,
with $q^2=M_V^2$.
 We have factored out the Born level cross section $\hat{\sigma}^B_{I}( \hat{s},\hat{t})$. The parton momenta $p_1^\mu$ and $p_2^\mu$
 carry momentum fractions $x_1$ and $x_2$ and are related to the hadron momenta via $p_1^\mu=x_1 P_1^\mu$ and $p_2^\mu=x_2 P_2^\mu$. The hadronic cross section is obtained after convoluting with parton distribution functions (PDFs) and summing over all partonic channels. 

At the partonic threshold the mass of the final-state jet 
$m_X^2=(p_1+p_2-q)^2$ vanishes. Since the full 
final state also involves the remnants of the scattered hadrons, this condition does not imply that the hadronic invariant mass $M_X^2=(P_1+P_2-q)^2$ vanishes, unless the momentum fractions $x_1$ and $x_2$ are close to unity.
Because there is no phase space for additional hard emissions near the threshold, the only corrections to the cross section arise from virtual corrections, encoded in the hard function $H_{I}(\hat{s},\hat{t},M_V,\mu)$, and soft and collinear emissions given by $S_{I}(k,\mu)$ and $J_{I}(p_J^2)$. The convolution over the soft momentum in (\ref{factform}) arises because the partonic jet mass is
\begin{equation}
m_X^2 = (p_J + k_S)^2 \approx   p_J^2 +2E_J k\,, 
\end{equation} 
where $p_J^\mu$ and $k_S^\mu$ are the collinear and soft momenta in the jet,
$E_J$ is the jet energy and $k=p_J\cdot k_S/E_J$.
 Since the reaction in the threshold region proceeds via Born-level kinematics, only two partonic channels are relevant: the Compton channel $q g \to q V$ and the annihilation channel $q\bar q \to g V$.

 A detailed derivation of the factorization formula (\ref{factform}) was given in \cite{Becher:2009th}. That paper focused on  photon production, but exactly the same jet and soft functions are relevant also for $W$ and $Z$ production. Explicit expressions for these functions in both the Compton and annihilation channel can be found in \cite{Becher:2009th}, together with the relevant anomalous dimensions. The non-zero boson mass only enters the hard function $H_{I}(\hat{s},\hat{t},M_V,\mu)$ and modifies the kinematics. The hard function is given by the virtual corrections to the corresponding hard scattering channel. To obtain the function at NLO, one needs the interference of the one-loop amplitude with the tree-level result, which is given in  equations (A.7) to (A.9) of \cite{Arnold:1988dp}. These expressions correspond to the bare result for the hard function. After renormalization, we obtain
\begin{multline}
 H_{q\bar q}(\hat{s},\hat{t},M_V,\mu) = 1+\frac{\alpha_s}{4\pi} \bigg[ \left(-C_A-2 C_F\right) \ln ^2\frac{\mu ^2}{\hat{s}} \\
 -2 \ln\frac{\mu ^2}{\hat{s}} \left(C_A \ln \frac{\hat{s}^2}{\hat{t} \hat{u}}+3 C_F\right) 
-6 C_F \ln\frac{\hat{s}}{M_V^2} \\
 -C_A \ln ^2\frac{\hat{t} \hat{u}}{\hat{s} M_V^2} -C_A \ln ^2\frac{\hat{s}}{M_V^2 } +C_F \left(\frac{7 \pi ^2}{3}-16\right)  \\
   + C_A \left(f(\hat{t})+f(\hat{u}) \right) \bigg] + 
 \Delta {\cal L}(\hat{s},\hat{t}) /\hat{\sigma}_{q\bar q}^{B}\,
 \end{multline}
 for the hard function in the annihilation channel, with 
 \begin{equation}
f(\hat{t}) = 2 \text{Li}_2\left(\frac{M_V^2}{M_V^2-\hat{t}}\right)+\ln^2\frac{M_V^2-\hat{t}}{M_V^2}+\frac{\pi ^2}{12} \,.
 \end{equation}
 The extra piece $\Delta {\cal L}(\hat{s},\hat{t})$ is the part of the virtual corrections not proportional to the Born cross section. It is finite and given in the last five lines of (A.9) of~\cite{Arnold:1988dp}.\footnote{The result is also given in \cite{Gonsalves:1989ar} but the sign of the $\Delta {\cal L}$ terms in expression (A4) in this reference is incorrect. We thank R.~Gonsalves for helping us to resolve this discrepancy.} The virtual corrections for the Compton channel are related to the above by crossing, see \cite{Arnold:1988dp} for the necessary relations.
 
  \begin{figure}[t!]
\begin{center}
\psfrag{m}[t]{$\mu$}
\psfrag{h}{$\mu_h$}
\psfrag{j}{$\mu_j$}
\psfrag{s}{$\mu_s$}
\psfrag{f}{$\mu_f$}
\psfrag{H}[]{$H_{I}(\hat{s},\hat{t})$}
\psfrag{J}[]{$J_{I}(m_X^2)$}
\psfrag{S}[]{$S_{I}(k)$}
\psfrag{F}[l]{$f_1(x_1)f_2(x_2)$}

\includegraphics[width=0.75\hsize]{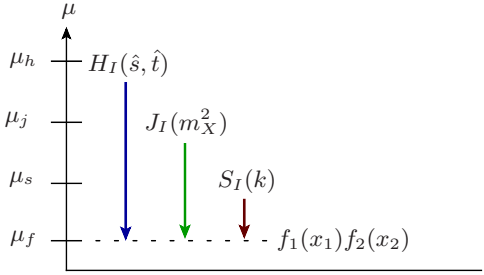}
\end{center}
\vspace{-0.5cm}
\caption{Resummation by RG evolution.\label{running}}
\end{figure}

Instead of integrating over the momentum fractions $x_1$ and $x_2$ of the two partons involved in the hard scattering, we follow \cite{Ellis:1981hk} and change variables to
\begin{equation} \label{kinematics}
\int _0^1 \rd x_1 \rd x_2 \theta(m_X^2) = 
\int_{x_{\rm min}}^1 \frac{\rd x_1 }{x_1 s+u- M_V^2 }\int_0^{m_{\rm max}^2} \rd m_X^2\,,
\end{equation} 
with $u=(P_2-q)^2=M_V^2-\sqrt{s}\sqrt{M_V^2+p_T^2}e^y$ and
\begin{align*}
m_{\rm max}^2&=u + x_1 (M_X^2- u) \, , & x_{\rm min}&=\frac{-u}{M_X^2-u}\,.
\end{align*}
In the above variables, the expansion around partonic threshold $m_X^2=0$ is performed at fixed $x_1$, which is the same as fixed $\hat{t} =(x_1 P_1-q)^2$. However, simply using (\ref{kinematics}) would be problematic, since the choice of expansion variables is not invariant under crossing $p_1\leftrightarrow p_2$, which would induce unphysical asymmetries in the rapidity spectrum. To avoid this, we symmetrize by performing the expansion around threshold twice: once at fixed $x_1$ using (\ref{kinematics}) and once at fixed $x_2$. Our result for the cross section is the average of the two expansions.

To perform the resummation of the soft and collinear emissions, the hard, jet and soft functions
must be evaluated at their own characteristic scales and then evolved to the factorization scale using the renormalization group (RG) where they are combined with the PDFs. This is illustrated in Figure~\ref{running}. 
The running is straightforward using the Laplace space formalism developed in \cite{Becher:2006nr}. 

\begin{figure}[t!]
\begin{center}
\includegraphics[height=0.56\hsize]{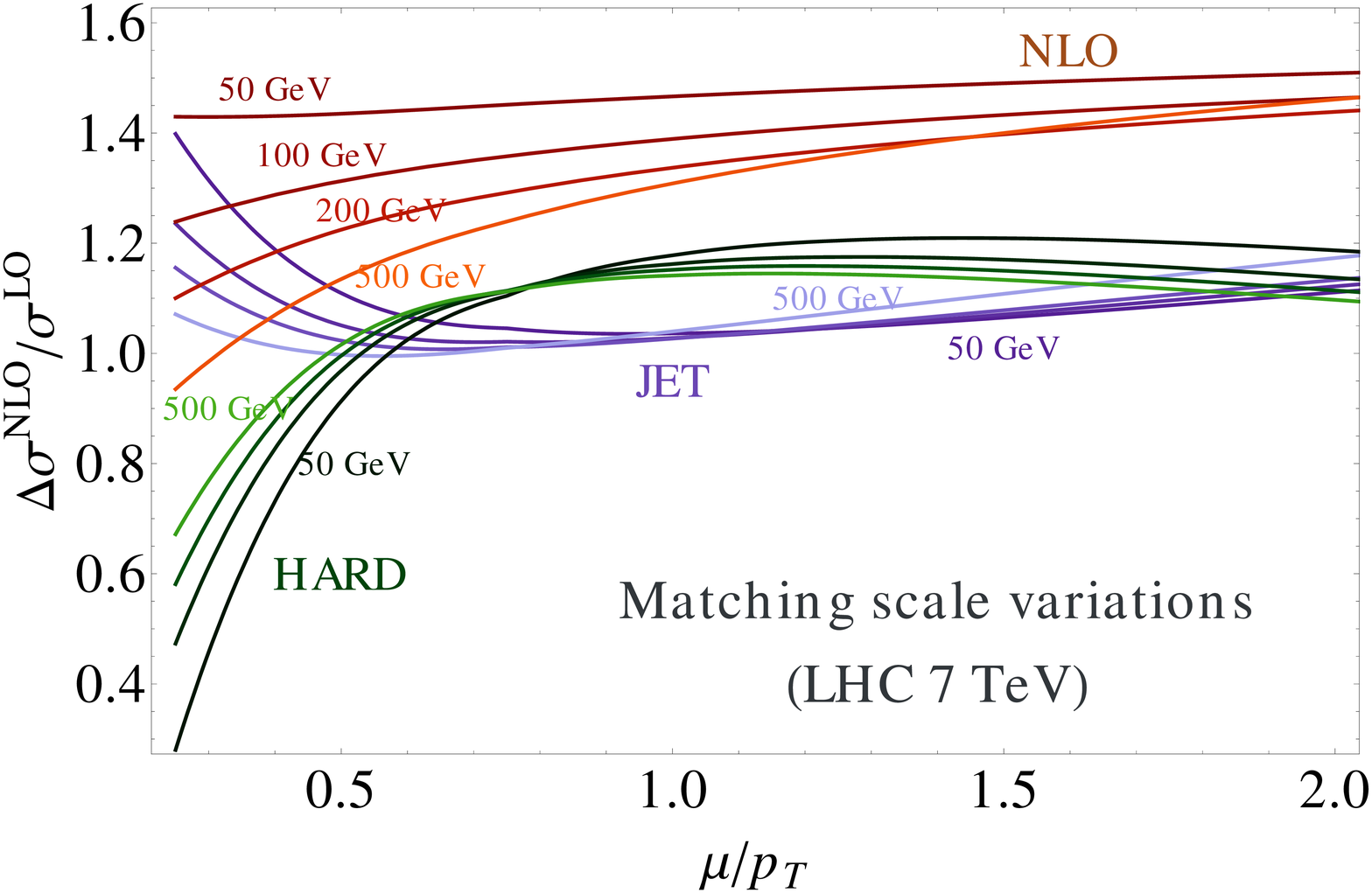} \\
\hspace*{0.15cm}\includegraphics[height=0.573\hsize]{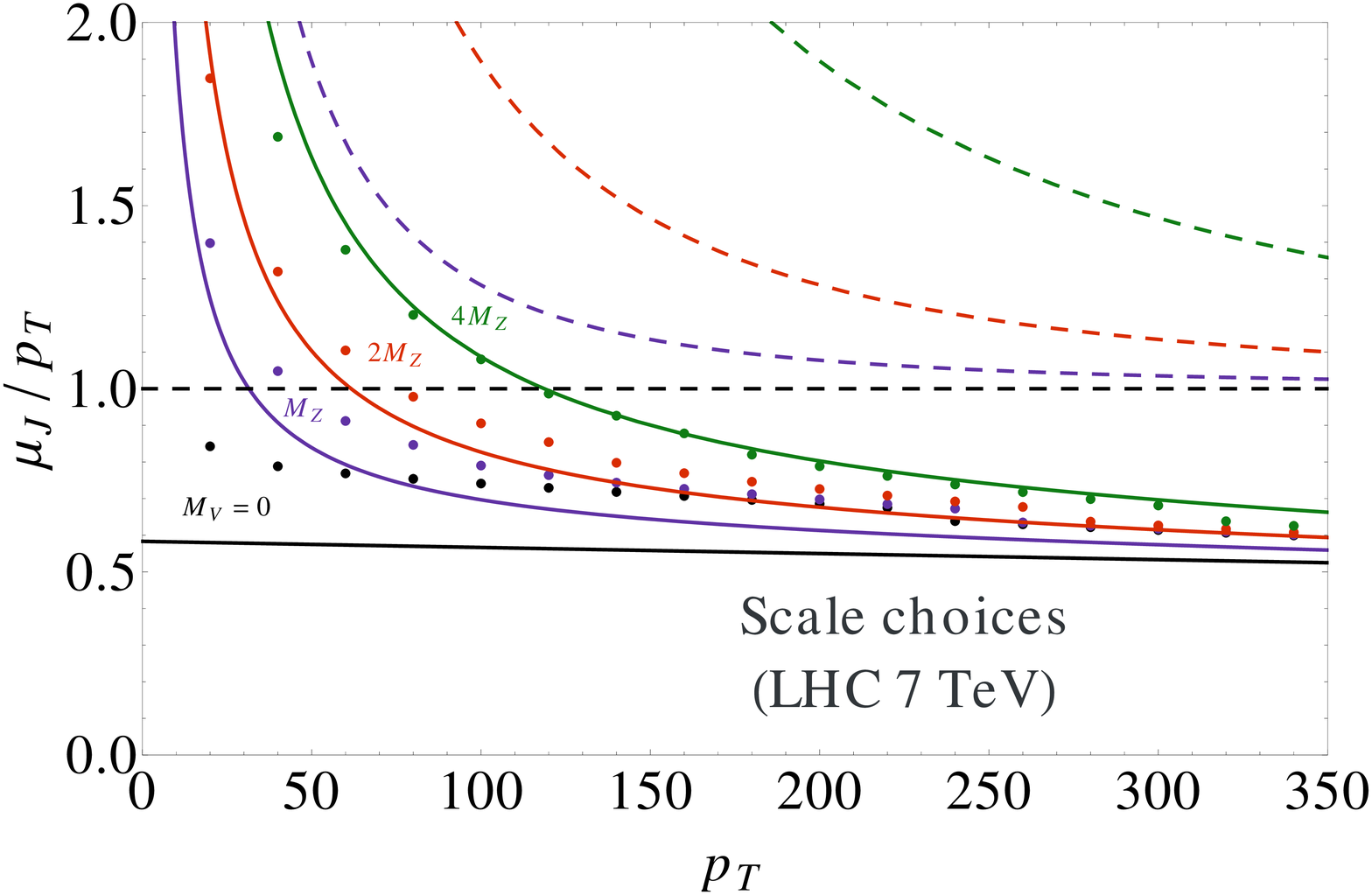}
\end{center}
\vspace{-0.5cm}
\caption{Scale setting. The plots are for $W^+$ bosons, but the qualitative features are the same for all bosons. \label{scalefig}}
\end{figure}

The most naive approach to scale setting would be to set the jet scale $\mu_j$ equal to the partonic jet mass $m_X$. However the partonic jet mass is not an observable; it is integrated over in the convolution with the PDFs. So setting $\mu_j=m_X$
one will encounter the Landau pole in the strong coupling constant, since the partonic jet mass $m_X$ can become arbitrarily small. The problematic choice $\mu_j=m_X$ is inherent in the traditional formalism for resummation. To avoid the Landau pole previous work did not perform the resummation to all orders and instead just computed the singular terms in the partonic cross section at the next, or the next two orders in perturbation theory~\cite{Kidonakis:1999ur,Kidonakis:2003xm}. In our work, we resum to all orders but choose $\mu_j$ to be the average-jet mass. Since the partonic cross section is convoluted with the PDFs,
the average jet mass must be calculated numerically.

To obtain the proper scale for each ingredient of the factorization formula  (\ref{factform}), we use the numerical procedure advocated in \cite{Becher:2007ty}. We evaluate the factorization theorem (\ref{factform}) numerically at a fixed scale $\mu$ and study individually the impact of the NLO corrections from the hard, jet and soft functions. The scale variations are shown on the
top panel in Figure~\ref{scalefig}. 
 Note that the jet and hard function variations have natural extrema. These extrema
are shown as the points in the lower panel for the jet scale. The solid curves are a reasonable approximation to these points, given by
\begin{equation}
\begin{aligned} \label{scalevalues}
\mu_h &=  \frac{13 p_T+ 2 M_V}{12} -\frac{p_T^2}{\sqrt{s}}\, ,\\
\mu_j &= \frac{7 p_T + 2 M_V }{12} \left (1-\frac{2p_T}{\sqrt{s}}\right) \, ,
\end{aligned}
\end{equation}
which we use instead of the exact extrema for simplicity. We also set $\mu_s=\mu_j^2/\mu_h$, as dictated by the factorization theorem. By choosing scales close to these extrema, we minimize the scale uncertainty. 

The scale setting procedure beautifully illustrates the power of the effective field theory approach. In a fixed-order computation, the hard, jet and soft corrections cannot be separated and are included at a common value of the renormalization scale. This is shown in the NLO curves in the top of Figure~\ref{scalefig}. In this case the scale dependence is monotonous. Because there are multiple relevant scales in the problem, there is no natural scale choice at fixed order. But there are
natural choices when $\mu$ is split into hard, jet and soft. For illustration, we also show in the bottom panel
of Figure~\ref{scalefig} the popular scale choice $\mu_j = \sqrt{p_T^2 + M_V^2}$, which is a bad fit to the jet scale.

The most natural choice for the factorization scale $\mu_f$ would be at or below $\mu_s$, since this scale defines the boundary between the perturbative and non-perturbative part of the process. However, since all PDF fits were performed with $\mu_f$ set equal to the hardest scale in the process, we will follow this convention and use  $\mu_f=\mu_h$ as our default value. Note that this implies that we use the RG to run the jet and soft functions from lower to higher values, in contrast to the situation depicted in Figure \ref{running}.

In order for our results to contain the full NLO cross section we match the resummed result to fixed order. The matching is straightforward since the resummation switches itself off when we set all scales equal, $\mu_h=\mu_j=\mu_s=\mu_f$. Doing so in the NNLL result yields all logarithmically enhanced terms at one loop level, which we denote by NLO$_{\rm sing}$. The matched result is obtained by adding the difference between the full NLO result and the singular terms NLO$_{\rm sing}$ to the NNLL resummed result. We denote the matched result by NNLL+NLO. To compute the NLO fixed-order result, we use the code {\sc qt}~\cite{qt}, and have verified that it agrees with {\sc mcfm}~\cite{mcfm}.

\begin{figure}[t!]
\begin{center}
\includegraphics[height=0.6\hsize]{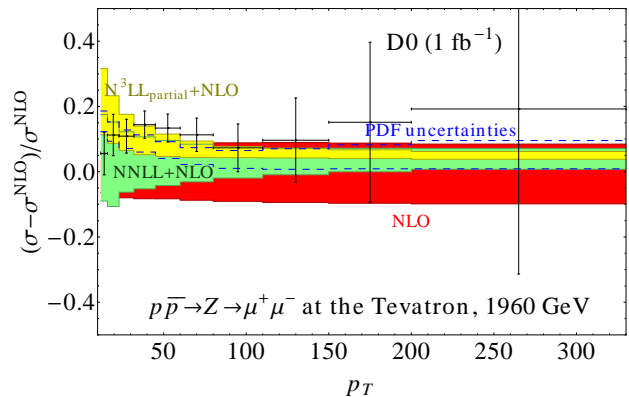}
\end{center}
\vspace{-0.5cm}
\caption{Comparison to DZero results \cite{Abazov:2010kn}.\label{tev}}
\end{figure}

\begin{figure}[t]
\begin{center}
\includegraphics[height=0.6\hsize]{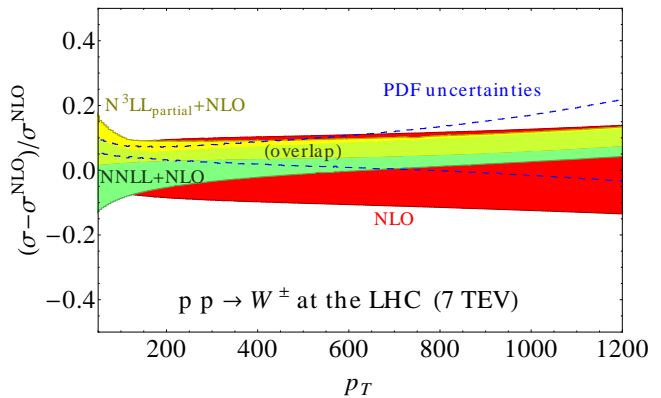}
\end{center}
\vspace{-0.5cm}
\caption{Prediction for the combined $W^+$ and $W^-$ cross sections the LHC (7 TeV).\label{lhc}}
\end{figure}

Most of the ingredients necessary to go to N$^3$LL accuracy are already known. The 3-loop anomalous dimensions
are all known. We use these, along with the two-loop jet function constants~\cite{Becher:2006qw,Becher:2010pd} and the Pad\'e
approximant for the 4-loop cusp anomalous dimension, to get our most accurate prediction, which we denote by N$^3$LL$_{\rm partial}$. 

For fixed order, we set the renormalization and factorization scales equal to $\mu_h$. 
 Uncertainties are estimated by varying by a factor of 2 around the defaults values \eqref{scalevalues} and extracting the maximum and minimum values.
For the final scale variation error bands on the resummed distributions,
 we  add the jet, hard, soft and factorization scale uncertainties in quadrature.
 All numerical predictions are computed using MSTW2008NNLO PDFs \cite{Martin:2009iq}, with $\alpha_s(M_Z)=0.1171$. For the electroweak parameters, we use $\alpha=1/127.92$, $\sin\theta_W=0.2263$, $M_W=80.40\, {\rm GeV}$, $M_Z=91.19\, {\rm GeV}$. 

In Figure \ref{tev}, we show the $p_T$ spectrum of the $Z$-boson at the Tevatron in comparison to results of the D0 experiment \cite{Abazov:2010kn}. Our results agree well with the experimental results but have significantly smaller uncertainties, in the region of high transverse momentum.

In Figure \ref{lhc}, we give results for the production of $W$ bosons at the LHC. The perturbative uncertainty on our result is much smaller than the PDF uncertainty, which implies that these results could be used to obtain more precise determinations of the PDFs, once the experimental results become available.  
\begin{figure}[t!]
\begin{center}
\includegraphics[height=0.55\hsize]{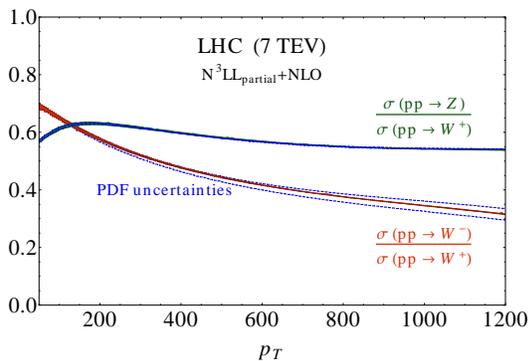}
\end{center}
\vspace{-0.5cm}
\caption{Prediction for ratios of cross sections for different gauge bosons at the LHC (7 TeV).\label{ratio}}
\end{figure}

In Figure \ref{ratio}, we give the ratio of the $W^-$ to $W^+$ cross section, as well as the ratio of $Z$'s to $W$'s. Most theoretical uncertainties drop out in these ratios, making them precision observables. Roughly half as many $W^-$ as $W^+$ bosons are produced at the LHC, which is just a reflection of the quark content of the protons, while the $Z$ to $W^\pm$ ratio is essentially given by the ratio of the electroweak couplings. Note that the PDF uncertainty is larger on the ratio involving the $W^-$, since the
$W^-$ is sensitive to sea-quark PDFs.

In Table~\ref{tab}, we give numerical results for integrated cross section with $p_T > 200$ GeV at the Tevatron and LHC.
In this table, we have included rows  NLO$_{\rm sing.}$ and NNLO$_{\rm sing.}$ which refer to the resummed distributions
expanded to fixed order. Indeed, Figure~\ref{scalefig} and the prescription (\ref{scalevalues}) make it clear that there is not a very large hierarchy between the different scales. For example, at the Tevatron, for $p_T=200\,{\rm GeV}$, the ratio between the scales is $\mu_h/\mu_s \approx 5$. Since the logarithms of the different scales are of moderate size, one can re-expand the resummed result in $\alpha_s$ and will find that most of the effect of the resummation is captured by the singular terms.
NLO$_{\rm sing.}$ refers to the expansion of the NNLL result (without matching) to $\alpha_s^2$.
NNLO$_{\rm sing.}$ refers to the expansion of the N$^3$LL$_{\rm partial}$ result to $\alpha_s^3$.
Comparing NLO to NLO$_{\rm sing.}$ we see that for $Z$-production with $p_T>200 {\rm GeV}$ at the Tevatron (LHC), 
we find that 82\% (70\%) of the NLO correction to the cross section is due to soft and collinear gluon emission.  Thus we
expect that SCET reproduces most of the NNLO result at high $p_T$.

Finally, there is one important effect that we have neglected, namely electroweak Sudakov logarithms. These can be large and must be included. Already at the Tevatron these are non-negligible and lower the cross section at $p_T=300\,{\rm GeV}$ by about $10\%$. At the LHC at 14TeV, the effect is about $-15\%$ at $p_T=500\,{\rm GeV}$, and around $-30\%$ for $p_T=2\,{\rm TeV}$~\cite{Kuhn:2004em,Hollik:2007sq}. To get a really accurate comparison to data, the electroweak corrections must be included.

\begin{table}[t!]
\begin{tabular}{lcccc}
& \multicolumn{2}{c}{Tevatron} & \multicolumn{2}{c}{LHC at $7\,$TeV} \\ 
& $W^\pm$ &  $Z$ & $W^\pm$ &  $Z$ \\\hline
LO & $0.91^{+0.19}_{-0.13}$ & $0.46^{+0.13}_{-0.09}$ & 
$34.5^{+4.6}_{-3.7}$ & $14.0^{+2.5}_{-2.0}$ \\ 
NLO$_{\rm sing.}$ & $1.19^{+0.07}_{-0.08}$ & $0.60^{+0.05}_{-0.06}$ & 
$47.1^{+2.1}_{-2.3}$ & $19.2^{+1.1}_{-1.2}$ \\ 
NLO & $1.28^{+0.09}_{-0.10}$ & $0.63^{+0.05}_{-0.06}$ & 
$53.3^{+3.8}_{-3.5}$ & $21.4^{+2.0}_{-1.9}$ \\ 
NNLL +NLO& $1.34^{+0.03}_{-0.03}$ & $0.66^{+0.02}_{-0.02}$ & 
$53.8^{+2.1}_{-2.1}$ & $22.5^{+1.0}_{-0.7}$ \\ 
NNLO$_{\rm sing.}$+NLO & $1.34^{+0.03}_{-0.04}$ & $0.65^{+0.02}_{-0.02}$ 
& $55.9^{+2.0}_{-1.4}$ & $21.7^{+1.1}_{-1.1}$ \\ 
N$^3$LL$_{\rm partial}$+NLO & $1.35^{+0.02}_{-0.02}$ & 
$0.66^{+0.01}_{-0.01}$ & $56.0^{+1.6}_{-0.7}$ & $22.6^{+0.7}_{-0.4}$ \end{tabular}
\vspace{-0.0cm}
\caption{Cross section $\sigma(p_T>200\,{\rm GeV})$ (in picobarn) using different approximations, see text\label{tab}. In addition to the scale uncertainties shown in the table, there is a relative PDF uncertainty of $3 \%$  ($2\%$) for $W^\pm$ production at the Tevatron (LHC) and $5 \%$  ($2\%$) for $Z$ production .}
\end{table}

It would be interesting to obtain NNLL predictions for less inclusive quantities. In particular, 
to compare to the preliminary CMS results for the $Z$ boson $p_T$ spectrum \cite{cmsres}, based on $36\,{\rm pb}^{-1}$, one would need results which are differential in the lepton momenta to account for the experimental cuts. To obtain these results, one has to compute the hard function for arbitrary boson polarizations, which is in progress \cite{christian}. Putting in cuts on the hadronic side is more complicated, but would be quite interesting, since it would allow to compute the production in association with a jet.

 The LHC has just delivered its first inverse femtobarn of data at 7 TeV.
 Comparing our results to this data will allow for precision tests of the standard model
at the highest energies ever produced in a collider.

{\em Acknowledgments:\/} The work of TB and CL is supported in part by the SNSF and ``Innovations- und Kooperationsprojekt C-13'' of SUK. TB and MDS would like to thank the Aspen Center for Physics 
and KITP and for hospitability during the completion of the project. This research was supported in part by the National Science Foundation under Grant No. NSF PHY05-51164 and by the Department of Energy, under grant DE-SC003916.


\begin{thebibliography}{1}
  
\bibitem{Ellis:1981hk}
  R.~K.~Ellis, G.~Martinelli, R.~Petronzio,
  Nucl.\ Phys.\  {\bf B211}, 106 (1983).

\bibitem{Arnold:1988dp}
  P.~B.~Arnold, M.~H.~Reno,
  Nucl.\ Phys.\  {\bf B319}, 37 (1989).

\bibitem{Gonsalves:1989ar}
  R.~J.~Gonsalves, J.~Pawlowski, C.~-F.~Wai,
  Phys.\ Rev.\  {\bf D40}, 2245 (1989).

\bibitem{Kidonakis:1999ur}
  N.~Kidonakis, V.~Del Duca,
  Phys.\ Lett.\  {\bf B480}, 87-96 (2000).
  
\bibitem{Kidonakis:2003xm}
  N.~Kidonakis, A.~Sabio Vera,
  JHEP {\bf 0402}, 027 (2004).
  
\bibitem{Gonsalves:2005ng}
  R.~J.~Gonsalves, N.~Kidonakis, A.~Sabio Vera,
  Phys.\ Rev.\ Lett.\  {\bf 95}, 222001 (2005).
  
  \bibitem{Bauer:2000yr}
 C.~W.~Bauer, S.~Fleming, D.~Pirjol and I.~W.~Stewart,
 Phys.\ Rev.\ D {\bf 63}, 114020 (2001).

\bibitem{Bauer:2001yt}
 C.~W.~Bauer, D.~Pirjol and I.~W.~Stewart,
 Phys.\ Rev.\ D {\bf 65}, 054022 (2002)

\bibitem{Beneke:2002ph}
 M.~Beneke, A.~P.~Chapovsky, M.~Diehl and T.~Feldmann,
 Nucl.\ Phys.\ B {\bf 643}, 431 (2002)
 
\bibitem{Becher:2009th}
  T.~Becher, M.~D.~Schwartz,
  JHEP {\bf 1002}, 040 (2010).

\bibitem{Becher:2006nr}
  T.~Becher, M.~Neubert,
  Phys.\ Rev.\ Lett.\  {\bf 97}, 082001 (2006).
  [hep-ph/0605050].

\bibitem{Becher:2007ty}
  T.~Becher, M.~Neubert, G.~Xu,
  JHEP {\bf 0807}, 030 (2008).

\bibitem{qt}
R.~Gonsalves, http://www.physics.buffalo.edu/gonsalves/.

\bibitem{mcfm}
J.~Campbell, K.~Ellis, C.~Williams, http://mcfm.fnal.gov/.
  

\bibitem{Becher:2006qw}
  T.~Becher, M.~Neubert,
  Phys.\ Lett.\  {\bf B637}, 251-259 (2006).
  [hep-ph/0603140].

\bibitem{Becher:2010pd}
  T.~Becher, G.~Bell,
  Phys.\ Lett.\  {\bf B695}, 252-258 (2011).


\bibitem{Martin:2009iq}
  A.~D.~Martin, W.~J.~Stirling, R.~S.~Thorne, G.~Watt,
  Eur.\ Phys.\ J.\  {\bf C63}, 189-285 (2009).

\bibitem{Abazov:2010kn}
  V.~M.~Abazov {\it et al.} [ D0 Collaboration ],
  Phys.\ Lett.\  {\bf B693}, 522-530 (2010).

  
\bibitem{cmsres}
CMS Collaboration, "Differential Cross Sections for Z Bosons at $\sqrt{s}=7$ TeV", CMS-PAS-EWK-10-010


\bibitem{Kuhn:2004em}
  J.~H.~Kuhn, A.~Kulesza, S.~Pozzorini, M.~Schulze,
  Phys.\ Lett.\  {\bf B609}, 277-285 (2005);
  Nucl.\ Phys.\  {\bf B727}, 368-394 (2005);
  Phys.\ Lett.\  {\bf B651}, 160-165 (2007);
  
  Nucl.\ Phys.\  {\bf B797}, 27-77 (2008).
  
\bibitem{Hollik:2007sq}
  W.~Hollik, T.~Kasprzik, B.~A.~Kniehl,
  Nucl.\ Phys.\  {\bf B790 } (2008)  138-159.

\bibitem{christian} Ch.\ Lorentzen, in progress.

  
\end{thebibliography}
\end{document}